# Analysis of recent *G* experiments by a differential version of MOND theory

Norbert Klein, Imperial College London, Department of Materials, London SW7 2AZ, UK

The discrepancy between recently reported experimental values of the gravitational constant *G* was analysed within a differential version of MOND theory. In contrast to the most commonly accepted interpretation of MOND theory, it is assumed that only the relative gravitational acceleration between a test mass and an array of source masses determines the magnitude of post Newtonian corrections at small magnitudes of acceleration. The analysis was applied to one of the most recent Cavendish-type experiments, which showed a significant deviation of the measured gravitational constant from the current CODATA value. A remarkable agreement between the observed *G* discrepancy and galaxy rotation curves was revealed by a consistent extrapolation within the framework of this model. The differential approach suggests that gravity - induced alterations of the space-time curvature may define the magnitude of corrections to Newton's law.

**Introduction**

More than 300 years after Newton, gravitation still remains one of the great miracles, and its understanding in the framework of a unified theory with the other fundamental forces is still lacking [1,2]. Newton's gravitational law, emerging as a first order solution for a point mass from Einstein's theory of general relativity (GR), allows highly accurate predictions of the motion of planets and other objects within our solar system. However, as first reported by Vera Rubin in 1970, the measured rotation of galaxies is incompatible with Newton's law, if only visible matter is considered for modelling the dynamic behaviour [3]. Therefore, the concept of dark matter was suggested as a possible explanation of this discrepancy, but dark matter has not yet been observed directly [4]. As an alternative to dark matter, the concept of Modified Newtonian Dynamics (MOND) was introduced by Milgrom in 1983 [5], which accounts for deviations from Newton's law in cases of small accelerations. Among the more recent successes of MOND theory the modelling of the dynamics of Andromeda dwarf galaxies [6] and gravitational lensing [7] are outstanding.

At the same time, accurate measurements of the gravitational constant *G* - most of them by Cavendish-type torsion balances [1,8,9], Fabry - Perot microwave resonators (direct involvement of the author) [10-12] and optical interferometers [13] suspended as linear pendulums are not in perfect agreement with each other, and the observed discrepancies of up to ten standard deviations remain unexplained to date [1]. Noticeably, the recently reported results by Quinn and Speake [8], as determined by two independent operation methods of a Cavendish-type *G* torsion balance, are 180 ppm above the results obtained by Schlamminger et al. [14], which are based on a beam balance employing 13 tons of liquid mercury as source mass – the latter being consistent with the most recent CODATA value [15]. As a possible explanation for these discrepancies, Anderson et al. suggested a sinusoidal time variation of *G* according to an analysis of *G* experiments being performed since 1980 [16], but according to a recent analysis of the consequence for orbital motions within our solar system by Iorio this scenario is in contradiction to the experimental constraints for the observed orbit increase of the LAGEOS satellite and anomalous perihelion precession of Saturn [17].

Inspired by the analysis of the acceleration magnitude dependence of gravitational force measurements by a pendulum gravimeter within MOND theory by Meyer et al. [12], in this report a post-Newtonian analysis of selected $G$ measurements is pursued and compared to galaxy rotation curves using different MOND extrapolation functions.

**A differential MOND version for the case of two point masses**

Two point masses $m_1$ and $m_2$ being separated by a distance $r_{12}$ are considered. In an idealized $G$ experiment, one of the two masses (called "test mass") is moved by an incremental distance from one equilibrium position (source mass at infinite distance from the test mass) to another (source mass at distance $r_{12}$ from the test mass) due to a small acceleration caused by the gravitational force $\vec{F}_{12}$ between $m_1$ and $m_2$. In a centre-of-mass frame of reference, the dynamics is described in relative coordinates, the inertia in Newton's second law is determined by the reduced mass $\mu$.

$$\vec{F}_{12N} = \mu \vec{a}_{rN} \qquad \vec{a}_{rN} = \vec{a}_{1N} - \vec{a}_{2N} \qquad \mu = \frac{m_1 m_2}{m_1 + m_2} \qquad \text{Eq. 1}$$

Although the two masses are arranged within a non-inertial frame of reference (the earth), it is presumed here that the post-Newtonian effect is solely determined by the gravitational force between $m_1$ and $m_2$: the test masses of a torsion pendulum represent a suitable approximation for a local inertial-frame of reference with respect for the direction in which the test masses move due to the gravitational force between test- and source masses. A possible scenario which underpins the presumption of this "differential" interpretation of MOND theory is presented in the discussion section.

In order to fit MOND-type post-Newtonian corrections at low magnitudes of acceleration, a universal extrapolation function $f(|\vec{a}_{rN}|)$

$$\vec{a}_r = \vec{a}_{rN} f(|\vec{a}_{rN}|); \qquad f(|\vec{a}_{rN}|) = \left[1 + \left(\frac{a_0}{|\vec{a}_{rN}|}\right)^\alpha\right]^{\frac{1}{2\alpha}} \qquad \text{Eq. 2}$$

is suggested, with $a_0 = 1.2 \cdot 10^{-10}$ m/s² resembling the fundamental acceleration parameter inherent to MOND theories [5], and $\alpha$ is a free parameter. $f(|\vec{a}_{rN}|)$ represents a correction of the relative Newtonian acceleration $\vec{a}_{rN}$ and becomes significant for small magnitudes of the latter. For the case of a negligible test mass ($m_1 \ll m_2$, $\mu = m_2$, $a_{rN} = a_N$) Eq. 2 fulfils the requirements for the asymptotic behaviour according to MOND theory, $a(a_N \gg a_0) \to a_N$ (Newtonian limit) and $a(a_N \ll a_0) \to (a_0 a_N)^{1/2}$ (deep MOND limit) [5]. The latter explains the Tully-Fisher relation [18], as one of the hallmarks of MOND theory.

The correction function according to Eq. 2 can be compared with solutions of the modified Poisson equation [19] for one point mass according to MOND theory for the most commonly used MOND extrapolation functions $\mu_{MOND0}(x) = [(1+4x)^{1/2} + ]/[(1+4x)^{1/2} - 1]$, $\mu_{MOND1}(x) = x/(1+x)$ and $\mu_{MOND2}(x) = x/(1+x^2)^{1/2}$ [5,12], which are

$$a_{MOND0} = a_N \left[ 1 + \sqrt{\frac{a_0}{a_N}} \right] \qquad \text{Eq. 3}$$

$$a_{MOND1} = a_N \left[ \frac{1}{2} + \sqrt{\frac{1}{4} + \frac{a_0}{a_N}} \right] \qquad \text{Eq. 4}$$

$$a_{MOND2} = a_N \sqrt{\frac{1}{2} + \frac{1}{2}\sqrt{1 + \left(\frac{2a_0}{a_N}\right)^2}} \qquad \text{Eq. 5}$$

It is worth noting that $a_{MOND0}$, which results from the relativistic generalization of MOND theory (TeVeS) [20], is identical to Eq. 2 for the parameter choice of $\alpha = \frac{1}{2}$. The practical advantage of the suggested extrapolation function (Eq.2) is that the smoothness of the transition from the Newtonian regime to the deep MOND regime can be varied continuously by the choice of the parameter $\alpha$.

Inserting Newton's law of gravity into Eq. 2 and using the definition for the reduced mass (Eq. 1), Eq. 2 turns out to be equivalent to a modified gravitational force law

$$\vec{F}_{12} = G m_1 m_2 \frac{\vec{r}_{12}}{|\vec{r}_{12}|^3} \left[ 1 + \left( \frac{a_0 |\vec{r}_{12}|^2}{G(m_1 + m_2)} \right)^\alpha \right]^{\frac{1}{2\alpha}} \qquad \text{Eq. 6}$$

$F_{12}$ is symmetric in $m_1$ and $m_2$, i.e. Newton's 1st law "actio = reactio" is inherently fulfilled. Eq. 6 emphasizes that the MOND-corrected gravitational force exhibits a non-linear dependence on the masses $m_1$ and $m_2$ by which the force is generated. It is worth to note that a similar but slightly different expression for the force between two point masses, which contains sum terms of $m_1$ and $m_2$, was reported recently as explicit expression for a subclass of MOND theories [21].

Due to the non-linear nature of the modified gravitational force, the superposition principle is violated, i.e. the effective force on a test mass $m_1$ by an ensemble of source masses $m_i$, $i = 2,…,N$ cannot be calculated by a simple vector superposition of the force as described by Eq. 6.

**Forces between a test mass and multiple field masses**

The post-Newtonian correction for an array of source masses, as often being used in $G$ experiments, is evaluated. We consider $k-1$ point-type source masses $m_2….m_k$, located at positions $r_i$, $i=2..k$, and the test mass $m_1$ located at $r_1$. Usually, in a $G$ experiment, only one component of this force is measured, determined by a unit vector $n$. In a Cavendish-type $G$ experiment, this component generates the measured torque (see appendix).

The $n$-component of the Newtonian acceleration of $m_1$ caused by the gravitational force between the array ($m_2…m_k$) and $m_1$ is given by

$$a_{1N} = G \sum_{i=2}^{k} \frac{m_i}{r_{1i}^2} \cos \varphi_i \qquad \text{Eq. 7}$$

with $\varphi_i$ representing the angle between $\mathbf{r}_{1i}=\mathbf{r}_i-\mathbf{r}_1$ and $\mathbf{n}$, and $r_{1i}$ the distance between $m_1$ to $m_i$. Since the source masses $m_2,…,m_k$ are rigidly connected, it is appropriate to calculate the $\mathbf{n}$-component of acceleration of the entire array $m_2,…,m_k$ due to the gravitational force between the array and $m_1$.

$$a_{2..kN} = -\frac{G}{\sum_{i=2}^{k} m_i} \sum_{i=2}^{k} \frac{m_1 m_i}{r_{1i}^2} \cos \varphi_i \qquad \text{Eq. 8}$$

According to Eqs. 1, 7 and 8 the post-Newtonian correction is determined by the relative acceleration

$$a_{rN} = a_{2..kN} - a_{1N} = G \left( 1 + \frac{m_1}{\sum_{i=2}^{k} m_i} \right) \sum_{i=2}^{k} \frac{m_i}{r_{1i}^2} \cos \varphi_i \ . \qquad \text{Eq. 9}$$

The term in parenthesis deviates from unity for the case that the value of the test mass ($m_1$) cannot be neglected in comparison to the source masses ($m_2,…,m_k$). Hence, the post-Newtonian correction is determined by Eq. 2, employing Eq. 9 for the evaluation of the Newtonian relative acceleration.

It is worth emphasizing that this approach to calculate forces due to multiple source masses is free of inconsistencies. For illustration, consider two identical source masses $m_2=m_3=M$ at the same distance $r_{12}=r_{13}=R$ from the test mass $m_1$. In this case $a_{rN} = G\left(1+\frac{m_1}{2M}\right)\frac{2M}{R^2}$ according to Eq. 9, which is identical to the result for a single test mass of value $2M$ at a distance $R$. Therefore, Eq. 2 in conjunction with Eq. 9 represents a reasonable and logical approach to deal with multiple source masses within MOND.

**Analysis of recent *G* experiments**

There are significant discrepancies between *G* values measured by different techniques, and *G* is therefore much less accurately determined in comparison to other fundamental constants [1]. As outstanding recent experiments, the work by Schlamminger et al. (referred as $G_0$) [14] and Quinn and Speake [8] represent considerable steps forward in terms of the achieved accuracy and reproducibility. Noticeably, their results differ significantly from each other by 180 ppm – corresponding to five standard deviations of the experimental uncertainty.

$$G_0 = 6.674\,252(122) \cdot 10^{-11} \frac{\text{m}^3}{\text{kgs}^2} \qquad G_{Quinn} = 6.67545(18) \cdot 10^{-11} \frac{\text{m}^3}{\text{kgs}^2}$$

This discrepancy will be analysed within the suggested differential MOND approach.

Schlamminger et al. used mercury source masses of 13,000 kg. Therefore, according to the criterion discussed before, this particular experiment is well within the deep Newtonian limit - within the claimed measurement error. We consider $G_0$ as a reference value for the gravitational constant, which is supported by the fact that $G_0$ lies well within the error limit of the current CODATA value of $(6.67384\pm0.0008)\cdot 10^{-11}$ m$^3$/kgs$^2$ [15]. Apparently, the value measured by Quinn and Speake is significantly higher.

In order to analyse the data within the suggested model, the geometry of the experiment by Quinn and Speake needs be analysed carefully. As explained in detail in the appendix, the ***n***-component of the relative acceleration $a_{rN}$ is $6.14\cdot 10^{-8}$ ms$^{-2}$ = $512\cdot a_0$.

In order to fit the $\alpha$ value of the extrapolation function to this result we employ

$$\frac{a_r}{a_{rN}} - 1 = \frac{G_{Quinn}}{G_0} - 1 = (1.93 \pm 0.45) \cdot 10^{-4} = \left[1 + \left(\frac{a_0}{a_{rN}}\right)^\alpha\right]^{\frac{2}{\alpha}} - 1 \qquad \text{Eq. 10}$$

to account for possible post-Newtonian corrections. Based on the experimental value $G_0$ and $G_{Quinn}$ and their uncertainties, $\alpha$ ranges between 1.20 and 1.26, the best fit is achieved for $\alpha$=1.23. Obviously the fitted value of $\alpha$ depends on the choice of $a_0$.

In Fig. 1, the relative deviation between Quinn's and Schlamminger's result for *G* is presented as a function of the relative differential Newtonian acceleration being employed in Quinn's Cavendish experiment, the latter normalized to Milgrom's MOND acceleration $a_0$=1.2·10$^{-10}$m/s$^2$. It is assumed that Schlamminger's result represents Newton's law. The area between the red lines corresponds to the extrapolated range of $\alpha$ values which are compatible with the analysis within the experimental error bars. For comparison, Eq. 2 is also plotted for a range of distinct values of $\alpha$ (full lines), along with MOND extrapolation functions according to Eqs. 3-5 (dashed lines).

As a direct comparison to recent astrophysical data, the purple dots represent individual resolved measurements along the rotation curves of nearly 100 spiral galaxies [23]. The original data in [23] are presented as ratio of the squares of the measured orbital velocity and the calculated "baryonic" velocity, as being calculated from the Newtonian gravitational acceleration by the baryonic (= visible stars and interstellar gas) mass of the galaxy. Since the centrifugal acceleration is proportional to gravitational acceleration, this ratio is equal to the ratio of measured acceleration and Newtonian acceleration. This enables a direct comparison with the *G* data.

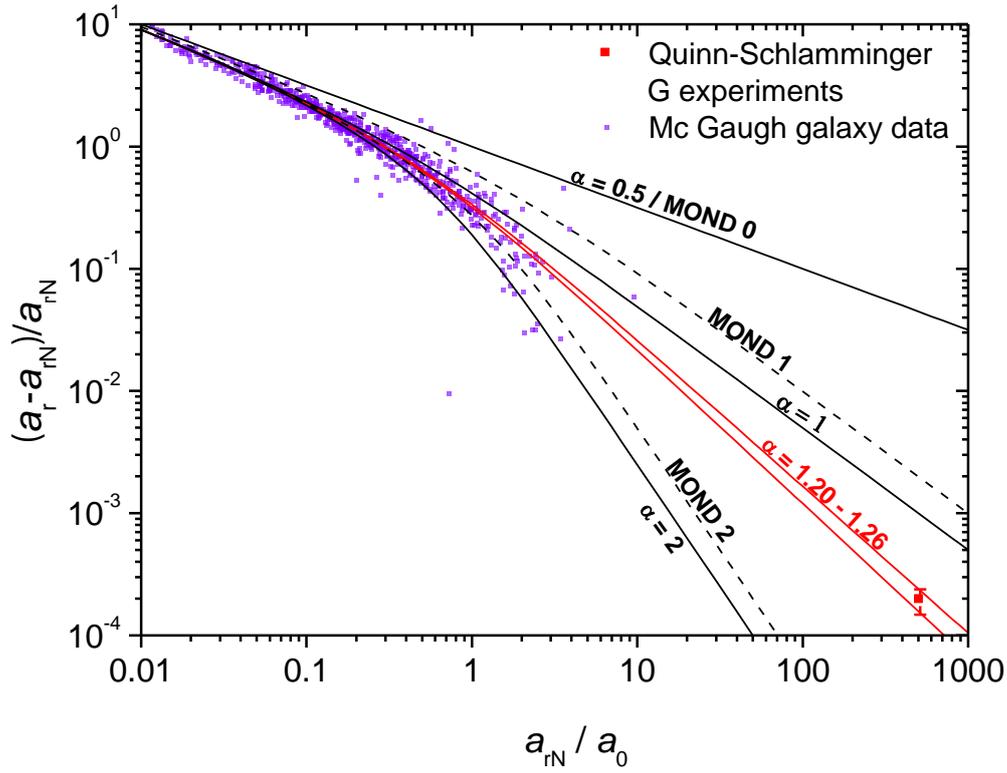

Fig.1: Comparison of the post-Newtonian acceleration determined from the Quinn-Schlamminger $G$ discrepancy with galaxy rotation data, in terms of relative deviation from Newton's law, $(a_r/a_{rN})-1$, as a function of the magnitude of the relative Newtonian acceleration $a_{rN}/a_0$. The experimental results are compared with a range of extrapolation functions according to MOND theories (see explanation in text). The red data point corresponds to the Quinn-Schlamminger $G$ discrepancy, the purple data points represent the galaxy data, and the area between the red curves represents the extrapolation function based on the suggested post-Newtonian model for the range of the parameter $\alpha$ being compatible with the Quinn-Schlamminger discrepancy within their error margins.

The extrapolation function extracted from the Quinn-Schlamminger discrepancy gives an excellent fit to the cloud of galaxy rotation curve data and is very close to MOND 2 for $a_{rN}/a_0 \leq 1$, which is the most relevant acceleration range for fits to galaxy rotation curves. Although the large scattering and measurement errors of the galaxy rotation data is compatible with a wider range of $\alpha$ values, it is worth to emphasize that the parameter $\alpha$ is solely determined from the terrestrial $G$-analysis, and not by a fit to the galaxy rotation data.

**Discussion**

Within a recent comprehensive review about $G$ experiments the authors concluded with the remark "The situation is disturbing—clearly either some strange influence is affecting most G measurements or, probably more likely, the measurements have unrecognized large systematic errors." [24] The presented analysis describes a possible scenario for "some strange influence", as a serious alternative to "unrecognized large systematic errors". On the other hand, among the $G$ experiments being published to date, the selection of Quinn's and Schlamminger's data is subjective, since other experiments show trends which are not compatible with the model presented here [1,9,13]. In particular, one of the more recent Cavendish-type experiment by Gundlach et al. [9], which operates in a similar acceleration range than the one by Quinn and Speake, has revealed a value close to that reported by Schlamminger et al. However, there is one important peculiarity in Gundlach's experiment: the field masses rotate at the extremely low rate of 20mrad /second on a radius of about 17 cm. In spite of this very slow rate, the corresponding centripetal acceleration is $a_c$=6.8·10$^{-5}$ m/s$^2$, which is 500,000 times larger than $a_0$. According to General Relativity (GR), gravity and acceleration are equivalent locally. In the framework of the differential interpretation of MOND theory, the difference of the tangential acceleration components of the test mass and the field masses needs to be taken into account in order to calculate the MOND correction, in spite that the acceleration of the field masses does not cause any torque. Since the direction of the centripetal acceleration changes during the rotation, one expects an additional change of the order of $a_c$, which needs to be considered for the calculation of the post-Newtonian correction according to Eq. 2. Therefore, for the given value of $a_c$ the post-Newtonian corrections for this experiment are expected to be less than 1 ppm according to the predictions by this model.

There is another important fact which may diminish the chances of observation of MOND effects in some terrestrial *G*-experiments: quite often, the amplitude of the periodic motion of the pendulum, driven by micro-seismic effects, is much bigger than the motion of the equilibrium pendulum position due to the gravitational force, in particular in case of a pair of linear pendulums [12,13], which exhibit stronger coupling to micro-seismic motion than torsion pendulums. Therefore, the micro-seismic-induced motion of each individual pendulum is likely to supress any MOND effect on a measurable scale, because the sum of the gravitational acceleration and the periodic acceleration of each pendulum has to be MOND-corrected. In case of the torsion experiment by Quinn and Speake the pendulum motion is either supressed (in cases of balancing the torque from the gravitational forces by an electrostatic torque, dubbed "servo method") or the amplitude of the oscillation is not bigger than the pendulum motion generated by the gravitational acceleration of the test mass (Cavendish method) [8]. The time-of-swing method at HUST (referred as HUST-09 [25]), which has resulted in a *G* value in agreement with the CODATA value, may not allow to observe MOND effects either due to a large amplitude of motion (not quoted in [25]) or due to the fact that the data acquisition time extends over a large number of pendulum periods of about 530 s each (in contrast to 120 s in case of the Quinn-Speake experiment): during the time the pendulum needs to change it equilibrium position after moving the source masses gravity acceleration component of moon and sun which points into the direction of the pendulum motion changes by an amount which is significantly larger than $a_0$. (for example, the acceleration of the moon may change by up to 1800 times $a_0$ over a time of 500 s, depending on the position of the moon, and the new equilibrium position mat be only after several pendulum oscillations). Although acceleration by moon and sun do not result in any measurable torque, they create an accelerated frame of reference for the test masses over the time span that the pendulum needs to get into its new equilibrium position, which may limit MOND effects to an unmeasurable magnitude.

The concept of a local inertial frame of reference with respect to the measurement direction appears more natural by looking at the effect of gravity from the perspective of the space time metric, which – according to GR – represents an equivalent description of gravity. In case of small masses such as those being used in terrestrial *G* experiments the space time curvature generated by an array of point masses can be described by the linearized Schwarzschild metric [26]

$$ds^2 = g_{00}c^2dt^2 + g_{jj}(dx^2 + dy^2 + dz^2) =$$
$$\left(1+\frac{2}{c^2}\Phi\right)c^2dt^2 - \left(1-\frac{2}{c^2}\Phi\right)(dx^2+dy^2+dz^2) \qquad j=1,2,3 \qquad \text{Eq. 11}$$

with *c* denoting the speed of light and Φ the gravitational potential by the source mass array $m_i$, i=1,…,k at locations $r_i$. In Eq. 11, the metric components $g_{jj}$ are determined by the gravitational potential of the source masses

$$\Phi(\vec{r}) = -G\sum_{i=2}^{k}\frac{m_i}{|\vec{r}-\vec{r}_i|} \qquad , \qquad \text{Eq.12}$$

for the case that the gravitational force by the test mass is neglected, i.e. $m_1 < m_i$ (i=2,..,k). According to Eqs. 11 and 12 the gravitational potential by the earth and large objects in the vicinity of the *G* experiment generate a nearly time and space independent contribution to the metric elements. In the context of the given MOND interpretation, post-Newtonian corrections may be related to the projection of the gradient of the space-time curvature onto the direction in which the gravitational force is being measured (unit vector **n**)

$$\vec{n}\vec{\nabla}\begin{Bmatrix}g_{00}(r)\\g_{jj}(r)\end{Bmatrix} = \pm\frac{2G}{c^2}\sum_{i=1}^{k}\frac{Gm_i\cos\varphi_i}{r_{1i}^2} \qquad , \qquad \text{Eq. 13}$$

which is equal to $2a_{rN}/c^2$ for the case of a small test mass.

An often overlooked evidence for a strong connection between MOND effects and space-time curvature arises from the numerical coincidence between the MOND acceleration parameter and the Hubble constant [5]

$$H_0 \approx 6\frac{a_0}{c} \qquad , \qquad \text{Eq.14}$$

which can be illustrated by re-visiting Einstein's famous elevator thought experiment [28]:

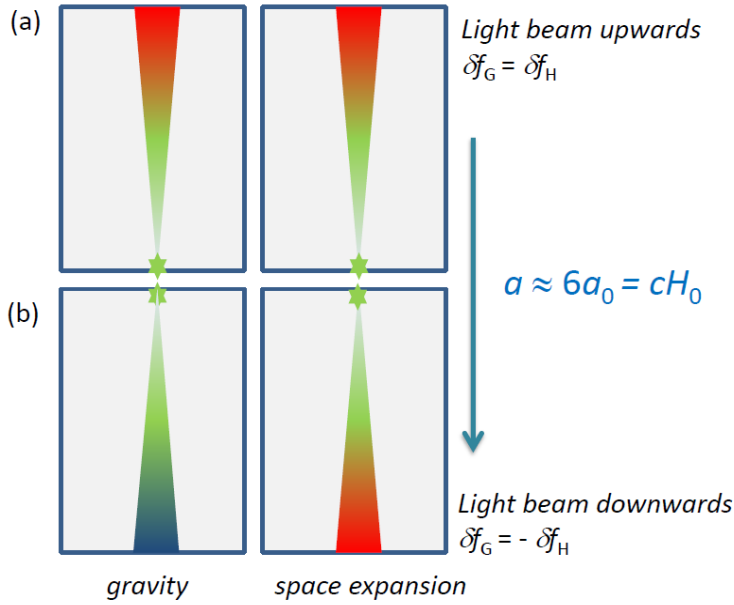

Fig.2: Illustration of Einstein's elevator thought experiment for the case of small accelerations of the order of the fundamental MOND acceleration $a_0$: due to the numerical coincidence with the Hubble constant $H_0$ gravitational redshift $\delta f_G$ and cosmological redshift $\delta f_H$ have the same magnitude at $a \approx 6a_0$ - in case of light source on the elevator floor (a). In case of source at the ceiling (b) $\delta f_G$ converts into a blue shift of same magnitude but $\delta f_H$ remains the same as in (a).

If the non-moving elevator is located in a homogeneous gravitational field, its Newtonian magnitude $a_N$ can be determined by an experimentalist inside the elevator via a simple red-shift experiment, as suggested by Einstein: Placing a highly monochromatic light source (green, see Fig. 2) of frequency $f_0$ on the floor of the elevator, and measuring the frequency $f_1$ by a detector placed at the ceiling of the elevator at a height $h$ above the floor, a simple relation between of redshift, $a_N$, and $h$ results according to GR.

$$\frac{\delta f_G}{f_0} \equiv \frac{f_1 - f_0}{f_0} = -\frac{a_N h}{c^2} \qquad \text{Eq. 15}$$

One of the key ingredient of Einstein's thought experiment is that Eq. 15 is identical to the Doppler shift which the experimentalist would observe if the elevator is located in an inertial frame of reference, but accelerated with $a_N$. In fact, historically the gravitational red shift was predicted in this way and later confirmed in the framework of the Schwarzschild metric [28].

In addition to the gravity induced redshift, a minuscule redshift occurs due to the cosmological expansion of the universe: the distance between two points in space in any direction increases with time at a minuscule rate - determined by the Hubble constant $H_0$. As a result of space expansion, the detector drifts from the light source at a speed $v=H_0 h$, which causes a small Doppler red shift. In order to separate the cosmological red shift from the gravitational shift, the experimentalist inside the elevator can swap light source and detector position and measure a gravitational blue shift (see Fig. 2 b). In his case he would still observe the same cosmological red shift as in the original experiment. Hence the total Doppler shift can be written as

$$\frac{\delta f}{f_0} = \frac{\delta f_H}{f_0} \pm \frac{\delta f_G}{f_0} = -\frac{H_0 h}{c} \mp \frac{a_N h}{c^2} \approx -\frac{h}{c^2} a_N \left( \frac{6a_0}{a_N} \pm 1 \right)$$ Eq. 16

with "plus" denoting the original configuration (light source on the floor) and "minus" the inverse configuration (light source at the ceiling). Due to the numerical coincidence between $a_0$ and $H_0/c$ the magnitude of the cosmological frequency shift $\delta f_H$ is identical to the gravitational red shift at $a_N \approx 6a_0$.

However, the amount of redshift at such small accelerations is minuscule at laboratory distances, for example $\delta f_G/f = 3.7 \cdot 10^{-36}$ for $h=1$ m. One may speculate that non-linear terms in the space-time metric result in combined terms containing both gravitational and cosmological redshift: as illustration, the simple quadrature of Eq. 16 generates a "±" term, which is proportional to $a_0 a_N$ (Eq. 17), resembling the deep MOND limit.

$$\left[ a_N \left( \frac{6a_0}{a_N} \pm 1 \right) \right]^2 = a_N^2 \left( 1 \pm \frac{12 a_0}{a_N} + \left( \frac{6a_0}{a_N} \right)^2 \right) = \pm 12 a_N a_0 + 36 a_0^2 + a_N^2$$ Eq. 17

As a proper solution of Einstein's field equations, the combined effects of weak gravitational forces and cosmological space expansion was formulated by McVittie in 1933 [29]. The Mc Vittie metric

$$ds^2 = \left[ \frac{1 - \frac{GM}{2rc^2 \beta(t)}}{1 + \frac{GM}{2rc^2 \beta(t)}} \right] c^2 dt^2 - \left[ 1 + \frac{GM}{2rc^2 \beta(t)} \right]^4 \beta^2(t)(dr^2 + r^2 d\Omega^2)$$ Eq. 18

represents a symbiosis between the Schwarzschild metric and the Robertson-Walker-Friedman-Lemaitre metric [28,29], the scaling factor $\beta(t)$ describes space expansion of a flat and isotropic universe. For short distances in comparison to the size of the universe $\beta(t)$ can be expressed by the Hubble constant $\beta(t)=\exp(H_0 t) \approx 1+H_0 t$. Hence, in the linear regime being relevant for weak forces the metric of a point mass inside an expanding universe can be simplified by neglecting terms smaller than $GM/rc^2$ and $H_0 t$ in Eq. 18.

$$ds^2 = \left( 1 - \frac{GM}{rc^2} \right) c^2 dt^2 - \left( 1 + 2H_0 t + \frac{2GM}{rc^2} \right)(dr^2 + r^2 d\Omega^2)$$ Eq.19

Similar to the previous discussion about the elevator experiment, it is apparent that the terms $H_0 t$ and $2GM/c^2$ describing tiny deviations of space components $g_{ii}$ from the Minkowski value $g_{11} = -1$, which are of same order of magnitude if the gravitational acceleration is of the order of $a_0$ (for $t \approx r/c$ = time which light needs to travel over a distance $r$ ).

Eq. 19 represents an elementary view on the connection between weak gravitational fields and cosmology based on existing concepts of GR, but it does not give any explanation for MOND effects. However, if modifications of Newton's law should be present at low accelerations, in particular, non-linear effects, which represent an inherent feature of MOND theory, Eq.19 illustrates that these modifications of gravity cannot be separated from cosmological space expansion. The differential interpretation of MOND seems logical, since external fields by planetary objects (as long as they do not move too much during the time scale of the pendulum oscillation) result only in $r$ and $t$ independent contributions to the metric elements – over the small length scale of the source mass motion in a *G* experiment. Therefore they do not have any physical relevance for the measured accelerations and forces, which are the only meaningful physical quantities in MOND theory. The gravitational field of the earth is perfectly compensated by the pendulum, and in the local reference system of the pendulum test mass it does not result in any *r* or *t*- dependent contribution to the space time metric - for the direction in which the gravitational force of the source masses is being measured.

The differential interpretation of MOND theory presented here is not consistent with MOND theory resulting from a modified Poisson equation [19] and its relativistic extension by Bekenstein [20]. In case of the modified Poisson equation the magnitude of the acceleration rather than one specific component appears as argument in the MOND extrapolation function, which make MOND effects unobservable in any terrestrial laboratory. As shown by Hees et al., MOND theories based on the modified Poisson equation can be ruled out from recent Cassini data on the orbital dynamics in our solar system for the range of extrapolation function being discussed here [30]. However, this analysis only refers to the specific effects of the modified Poisson equation and not to the modification of distance dependence of the gravitational force at low magnitudes as being discussed here.

Finally, the differential interpretation of MOND theory should not be confused with the inertia interpretation of MOND theory [5]. According to the differential interpretation, the MOND corrections do not apply to electromagnetic forces, in contrast to a general violation of second Newton's law within the inertia interpretation. Deviations from second Newton's law for electrostatic forces have been ruled out experimentally with precision far beyond the magnitude of $a_0$ [31].

**Conclusion**

The presented analysis has revealed a first indication that the observed discrepancies between *G*-values determined by different terrestrial experiments may have the same physical origin as the anomalies of galaxy rotation curves. Since unknown or underestimated systematic errors in current *G* experiment cannot be ruled out, this finding may just be a coincidence. In order to test this hypothesis, accurate measurements of small forces (gravitational and non-gravitational) in different regimes of acceleration by different methods are pivotal. It is recommended to run Cavendish experiments with a variety of torque magnitudes– either by comparison of different source masses or by using different angular positions for a given source mass array. Measurements of torque ratios may be less sensitive to some systematic errors than absolute *G* measurements. The dynamics of the experimental procedure (magnitude of pendulum oscillation, pendulum period, source mass movement, position of moon and sun during data acquisition) may have an influence on the results due to the subtle non-linear nature of MOND forces. Linear pendulum experiments seem to be less suited than Cavendish experiments, because they are more prone to strong and uncontrolled pendulum motions driven by micro-seismic activities. The presented model provides a tool and a guideline for data analysis within the differential version of MOND theory.

**Acknowledgements:**

The author likes to express his thanks to Clive Speake, University of Birmingham, UK, for checking the consistency of the analysis of his G experiment, Stacy Mc Gaugh from Case Western Reserve University for providing the galaxy data and Hinrich Meyer and Eberhard Wuensch, DESY, Germany, for motivating discussions.

**Appendix:**

In Quinn's experiment [8], the tangential component of the gravitational force between four field masses of $m_i$ = 11 kg ($i$=2,3,4,5) each, which are arranged on a circle of radius $R_2$ = 214 mm, and four test masses of $m_1$ = 1.2 kg each, arranged on a smaller concentric circle of radius of $R_1$ = 120 mm is determined. The circle of field masses is rotated by an angle $\varphi_0$ = 18.9° with respect to the circle of test masses, in order to maximize the torque. Although the field masses are short cylinders in Quinn's experiment, the point mass approximation is valid within an accuracy of a few percent [32], which is sufficient for the discussion of post-Newtonian corrections.

In Fig. A1, the tangential direction of force, which leads to the measured torque, is indicated by the unit vector **n**. In the following, the relative acceleration according to Eq. 9 caused by the gravitational forces of the four identical field masses on one arbitrary chosen test mass (the lower one highlighted in the Fig. A1) will be calculated.

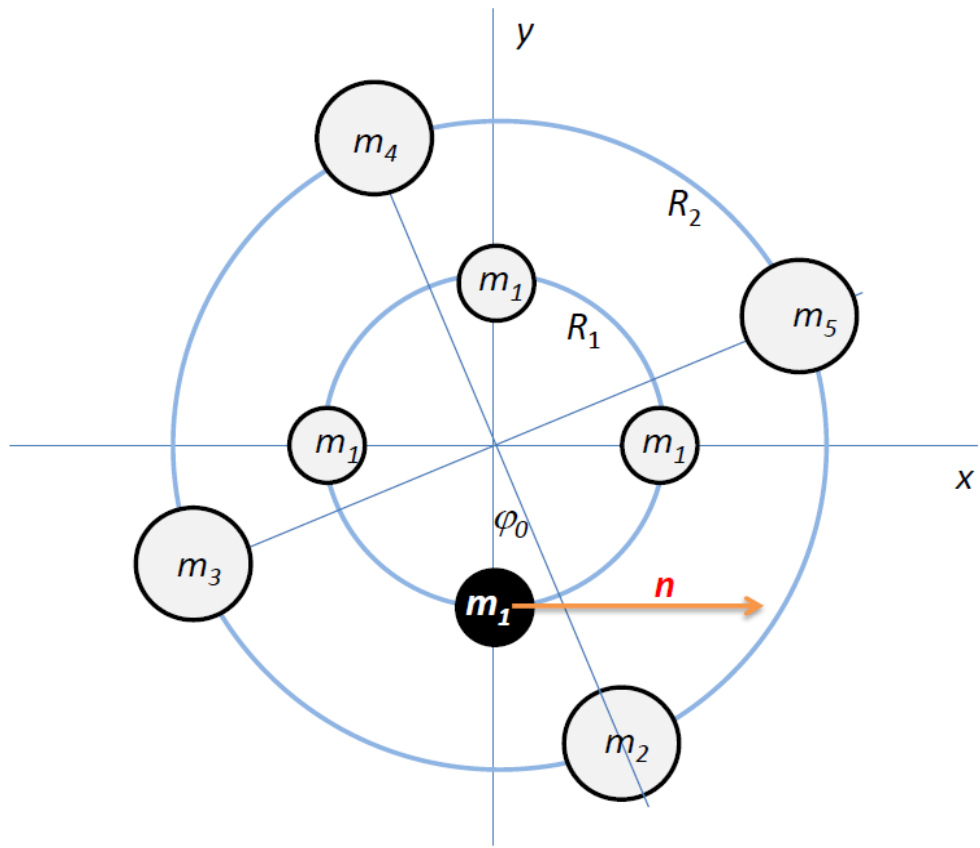

Fig. A1: Schematics of the Cavendish-type *G* experiment by Quinn et al. according to [5]

At first, the positions of the selected test mass and the four field masses are described as vector in Cartesian coordinates according to Fig. A1:

$$\vec{r}_1 = \begin{bmatrix} 0 \\ -R_1 \end{bmatrix}$$

$$\vec{r}_2 = R_2 \begin{bmatrix} \sin\varphi_0 \\ -\cos\varphi_0 \end{bmatrix}; \quad \vec{r}_3 = R_2 \begin{bmatrix} -\cos\varphi_0 \\ -\sin\varphi_0 \end{bmatrix}$$

$$\vec{r}_4 = R_2 \begin{bmatrix} -\sin\varphi_0 \\ \cos\varphi_0 \end{bmatrix}; \quad \vec{r}_5 = R_2 \begin{bmatrix} \cos\varphi_0 \\ \sin\varphi_0 \end{bmatrix}$$

Eq. A1

The distances $r_{1i}$ (i=2,..,5) between the highlighted test mass and field mass $m_i$, i = 2...5 are given by

$$r_{1i} = |\vec{r}_i - \vec{r}_1|$$

Eq. A2

yielding:

$$r_{12} = \sqrt{R_2^2 \sin^2\varphi_0 + (-R_2 \cos\varphi_0 + R_1)^2} = 107.7 \text{ mm}$$

$$r_{13} = \sqrt{R_2^2 \cos^2\varphi_0 + (-R_2 \sin\varphi_0 + R_1)^2} = 208.7 \text{ mm}$$

$$r_{14} = \sqrt{R_2^2 \sin^2\varphi_0 + (R_2 \cos\varphi_0 + R_1)^2} = 329.8 \text{ mm}$$

$$r_{15} = \sqrt{R_2^2 \cos^2\varphi_0 + (R_2 \sin\varphi_0 + R_1)^2} = 277.2 \text{ mm}$$

Eq. A3

As a next step, the values of $\cos\varphi_i$ (i=2...5), representing the angle between the vector $r_i$- $r_1$ and the tangential unit vector $n$ = (1,0) is calculated

$$\cos\varphi_i = \frac{(\vec{r}_i - \vec{r}_1)\vec{n}}{r_{1i}}$$

Eq. A4

yielding

$$\cos\varphi_1 = \frac{R_2 \sin\varphi_0}{r_{tf1}} = 0.643$$

$$\cos\varphi_2 = \frac{-R_2 \cos\varphi_0}{r_{tf2}} = -0.970$$

$$\cos\varphi_3 = \frac{-R_2 \sin\varphi_0}{r_{tf3}} = -0.267$$

$$\cos\varphi_4 = \frac{R_2 \cos\varphi_0}{r_{tf2}} = +0.743$$

Eq.A5

The different signs of the cosines refer to the obvious direction of each force component (see Fig. A1).

Inserting the numerical values from Eqs. A3 and A5 into Eq. 9 and using Schlamminger's value for $G$ yield a Newtonian relative acceleration $a_{rN}$=3.07·10$^{-8}$ m/s$^2$. Since the tray table with the source masses is moved between $-\varphi_0$ and $+\varphi_0$, during one run of the experiment, the relevant acceleration is twice as large, i.e. $6.14 \cdot 10^{-8}$ m/s$^2$.